\documentclass[%
superscriptaddress,
reprint,
showpacs,
 amsmath,amssymb,
 aps,
pra,
xcolor
]{revtex4-1}

\usepackage{color}
\usepackage{graphicx}
\usepackage{dcolumn}
\usepackage{bm}
\usepackage{amssymb}

\begin{document}

\title{
Ultralong spin coherence times for rubidium atoms in solid parahydrogen via dynamical decoupling
}

\author{Sunil Upadhyay}
\affiliation{Department of Physics, University of Nevada, Reno NV 89557, USA}
\author{Ugne Dargyte}
\affiliation{Department of Physics, University of Nevada, Reno NV 89557, USA}
\author{David Patterson}
\affiliation{Broida Hall, University of California, Santa Barbara, Santa Barbara, California 93106, USA}
\author{Jonathan D. Weinstein}
\email{weinstein@physics.unr.edu}
\homepage{http://www.weinsteinlab.org}
\affiliation{Department of Physics, University of Nevada, Reno NV 89557, USA}


\begin{abstract}
Coherence time is an essential parameter for quantum  sensing, quantum information, and quantum computation. In this work, we demonstrate electron spin coherence times  {\color{black} as long as} 0.1~s for an ensemble of rubidium atoms trapped in a solid parahydrogen matrix. 
We explore the underlying physics limiting the coherence time. The properties of these matrix isolated atoms are very promising for future applications, including quantum sensing of nuclear spins. If combined with efficient single-atom readout, this would enable NMR and MRI of single molecules co-trapped with alkali-metal atom quantum sensors within a parahydrogen matrix. 
\end{abstract}

\maketitle

Optical pumping and  detection of the spin states of ensembles of alkali atoms trapped in solid parahydrogen has been previously demonstrated \cite{upadhyay2016longitudinal}. The atoms exhibit excellent ensemble transverse spin relaxation properties, with a long spin dephasing time T$_2^*$  \cite{PhysRevB.100.024106, PhysRevA.100.063419}. In this work, we use Hahn spin echo \cite{PhysRev.80.580, PhysRev.94.630} and alternating-phase Carr-Purcell pulse sequences \cite{slichter1990} to measure the spin decoherence time T$_2$. 
We achieve a T$_2$ orders of magnitude longer than T$_2^*$. This --- combined with the localization possible through trapping in a solid matrix --- is very promising for applications in quantum sensing, nanoscale AC magnetometry \cite{PhysRevB.86.045214, barry2019sensitivity, RevModPhys.89.035002}, NMR of single molecules \cite{taylor2008high}, and nano-MRI \cite{PhysRevX.5.011001, staudacher2013nuclear, mamin2013nanoscale, sushkov2014magnetic}. 

We trap alkali-metal atoms in a solid parahydrogen matrix at a temperature of $\sim 3$~K, as described in references \cite{PhysRevA.100.063419, upadhyay2016longitudinal}. The samples are grown by vacuum deposition onto a cryogenic sapphire window. The vast majority of data we present is for rubidum atoms, due to their favorable properties as detailed in Ref. \cite{PhysRevA.100.063419}.
We can vary the alkali-metal atom and orthohydrogen density in the matrix. {\color{black} The alkali-metal atom density is measured by optical spectroscopy of the sample, and the orthohydrogen density is measured by Fourier-transform infrared spectroscopy \cite{upadhyay2016longitudinal}.} We typically work at total alkali-metal atom densities from $10^{16}$ to $10^{18}$~cm$^{-3}$. Typical sample thicknesses are on the order of 0.4~mm, and our pump and probe lasers select a volume of roughly $10^{-5}$~cm$^3$.  The number of alkali-metal atoms that we optically address is typically two orders of magnitude lower than the total atom number within that volume, as  reported in Ref. \cite{PhysRevA.100.063419}.

The degeneracy of the $m_\mathrm{F}$ levels is split by a magnetic bias field along the $\hat{z}$-axis. 
We optically pump the spin state of the implanted atoms with a pulse of high-intensity circularly-polarized light; we continuously measure their spin state through circular dichroism measurements \cite{PhysRevA.100.063419}. The pump and probe beams are nearly parallel to the $\hat{z}$-axis. 

We drive transitions between $m_\mathrm{F}$ levels with pulses of RF magnetic fields along the $\hat{y}$ axis, generated by an arbitrary waveform generator. We typically work at bias fields from $\sim10$ to $100$ Gauss; the higher end of that field range is sufficient to spectroscopically resolve the different $m_\mathrm{F}$ transitions due to the nonlinearity of the Zeeman effect \cite{PhysRevB.100.024106}. This allows us to isolate pairs of $m_F$ levels to create an effective two-level system out of the multi-level Zeeman-hyperfine structure.

\begin{figure}[ht]
    \begin{center}
    \includegraphics[width=\linewidth]{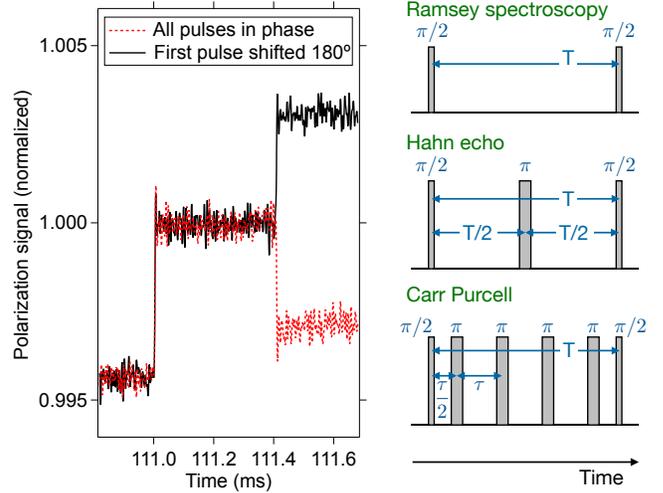}
    \caption{ 
    At the right, a schematic of pulse sequences. 
    At the left, typical data, as described  in the text. A initial $\pi/2$-pulse of the Ramsey sequence  at 111.0~ms creates a superposition and the final $\pi/2$-pulse at 111.4~ms provides readout. 
    In between, a sequence of pulses is applied for dynamical decoupling. Here, a single $\pi$-pulse was applied at 111.2~ms for a Hahn echo sequence.
\label{fig:RawDataAndPulseSequences}
    }
    \end{center}
\end{figure}

A typical data sequence is shown in Fig. \ref{fig:RawDataAndPulseSequences}. In this data, we apply the optical pumping beam from 52 to 102~ms (prior to the time window shown in Fig. \ref{fig:RawDataAndPulseSequences}). For $^{85}$Rb, this maximizes population in the $F=3$, $m_F = -3$  state. A sequence of RF ``pre-sweeps'' from  103 to 111~ms transfer population from $m_F=-3$ into the $m_F= -1$ level and reduces the residual population in the $m_F = 0$ level \cite{PhysRevB.100.024106}. We then perform Ramsey interferometry \cite{PhysRev.78.695} on the $F=3$, $m_F=-1 \leftrightarrow 0$ transition with two single-frequency $\pi/2$ pulses. We apply additional pulses between the two Ramsey $\pi/2$ pulses for dynamical decoupling, as detailed in Fig. \ref{fig:RawDataAndPulseSequences}.

The polarization signal shown in Fig. \ref{fig:RawDataAndPulseSequences} is the ratio of transmission of left-hand-circular and right-hand-circular probe beams through our sample, normalized to a level of 1 immediately before the readout pulse. We repeat the sequence with the phase of the first pulse shifted by 180$^\circ$, but otherwise unaltered. 
We take our signal amplitude to be the difference in the polarization signal between the two sequences after the final $\pi/2$ pulse.
This ensures the measured signal amplitude is due to coherence that has been maintained from the first Ramsey $\pi/2$ pulse and not an artifact from imperfections in our sequence.

\begin{figure}[ht]
    \begin{center}
    \includegraphics[width=\linewidth]{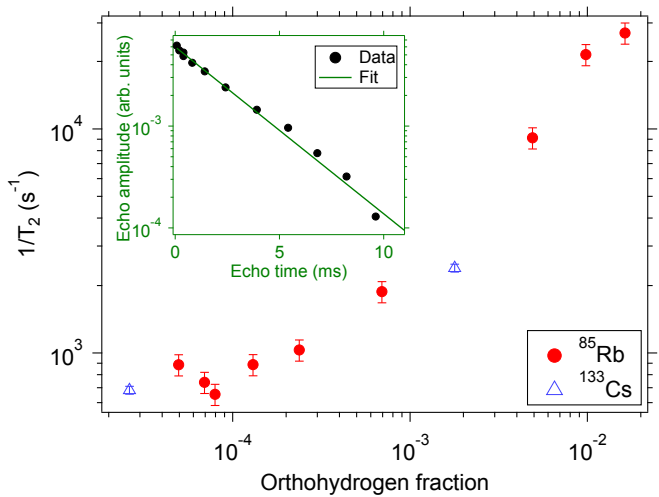}
    \caption{ 
The inset shows Hahn echo data for $^{85}$Rb, for a superposition of the $m_F = -1$ and $m_F = 0$ levels, fit to exponential decay.
%
%
The main graph shows the measured  decoherence rates ($1/\mathrm{T}_2$) for different samples, taken by traditional Hahn echo  with unresolved $m_F$ transitions, as described in the text. 
Rb data taken at a magnetic field of 13~G; Cs data taken at 22~G.  The orthohydrogen fraction is typically known to within $\pm 25\%$.
\label{fig:HahnEchoLifetimeOrthoCombo}
    }
    \end{center}
\end{figure}

We take Hahn echo data by the method described above, using the pulse sequence shown in Fig. \ref{fig:RawDataAndPulseSequences}. 
We measure the signal amplitude as a function of the delay between the first and final $\pi/2$ pulse. The resulting data is shown in Fig. \ref{fig:HahnEchoLifetimeOrthoCombo}. We fit this data to an exponential to extract the Hahn echo transverse relaxation time T$_2$. The values of T$_2$ obtained by this method are identical --- to within our signal-to-noise --- to the values obtained by doing traditional Hahn echo sequences  \cite{PhysRev.80.580} on unresolved $m_F$ levels at lower magnetic fields.

As seen in Fig. \ref{fig:HahnEchoLifetimeOrthoCombo}, the Hahn-echo spin coherence time depends critically on the fraction of orthohydrogen in our parahydrogen matrix. 
At  orthohydrogen fractions $\gtrsim 10^{-3}$, the decoherence rate  $(1/\mathrm{T}_2)$ increases  linearly in the orthohydrogen fraction. This is a clear indication that, at such concentrations, interactions with orthohydrogen are the dominant decoherence mechanism. %
Unlike parahydrogen molecules, which have nuclear spin $I=0$, orthohydrogen molecules have a nonzero magnetic moment with $I=1$. 
Due to their short T$_1$ \cite{buzerak1977nmr, washburn1983pulsed} and T$_2^*$ \cite{reif1953nuclear}, their nuclear spins would be expected to generate a stochastic fluctuating magnetic field. 
We note that decoherence from nuclear spins has previously been observed with NV centers in diamond \cite{childress2006coherent}, phosphorus donors in silicon \cite{PhysRevB.82.121201}, and other systems \cite{PhysRevB.90.241203}.
The nuclear spin purities reported here are not quite as good as what has been reported for  isotopically-purified diamond \cite{teraji2013effective, barry2019sensitivity}; we expect to achieve  lower nuclear spin densities in future work with modifications to our cryostat.

We note that  we observe no significant dependence of the Hahn-echo T$_2$ on the magnetic field over the range explored. Similarly, we observe no significant dependence on the alkali-metal atom density for densities  $\lesssim 10^{17}$~cm$^{-3}$. The longest measured Hahn-echo T$_2$ times of 2~ms  are significantly shorter than the longitudinal relaxation time T$_1$. {\color{black} Under typical probe beam conditions used in this paper we observe a T$_1$ of $\sim 0.2$~s,  reduced from the  $\gtrsim 1$~s T$_1$'s observed with the probe beam at a lower duty cycle \cite{upadhyay2016longitudinal, PhysRevA.100.063419}.
The probe's intensity and duty cycle are varied to verify that  it has a negligible effect on T$_2$.
}

At orthohydrogen fractions $\lesssim 10^{-4}$, the coherence time has little dependence on the ortho fraction.
We  probe the underlying physics limiting T$_2$ by comparing different species and different $m_F$ superposition states.
To compare different species, we  grow samples doped with $^{133}$Cs and (separately) with Rb atoms; the different $g$-factors of $^{85}$Rb and $^{87}$Rb  allow us to measure the two isotopes separately.
We measure the spin-echo T$_2$ for both $^{85}$Rb and $^{87}$Rb, following the protocol outlined above and in Fig. \ref{fig:RawDataAndPulseSequences}.
To compare different superposition states within the same species, we use single-photon RF transitions to produce superposition states of $m_F = 0$ and $m_F = -1$, and two-photon RF transitions to produce superposition states of $m_F = +1$ and $m_F=-1$.

\begin{figure}[ht]
    \begin{center}
    \includegraphics[width=\linewidth]{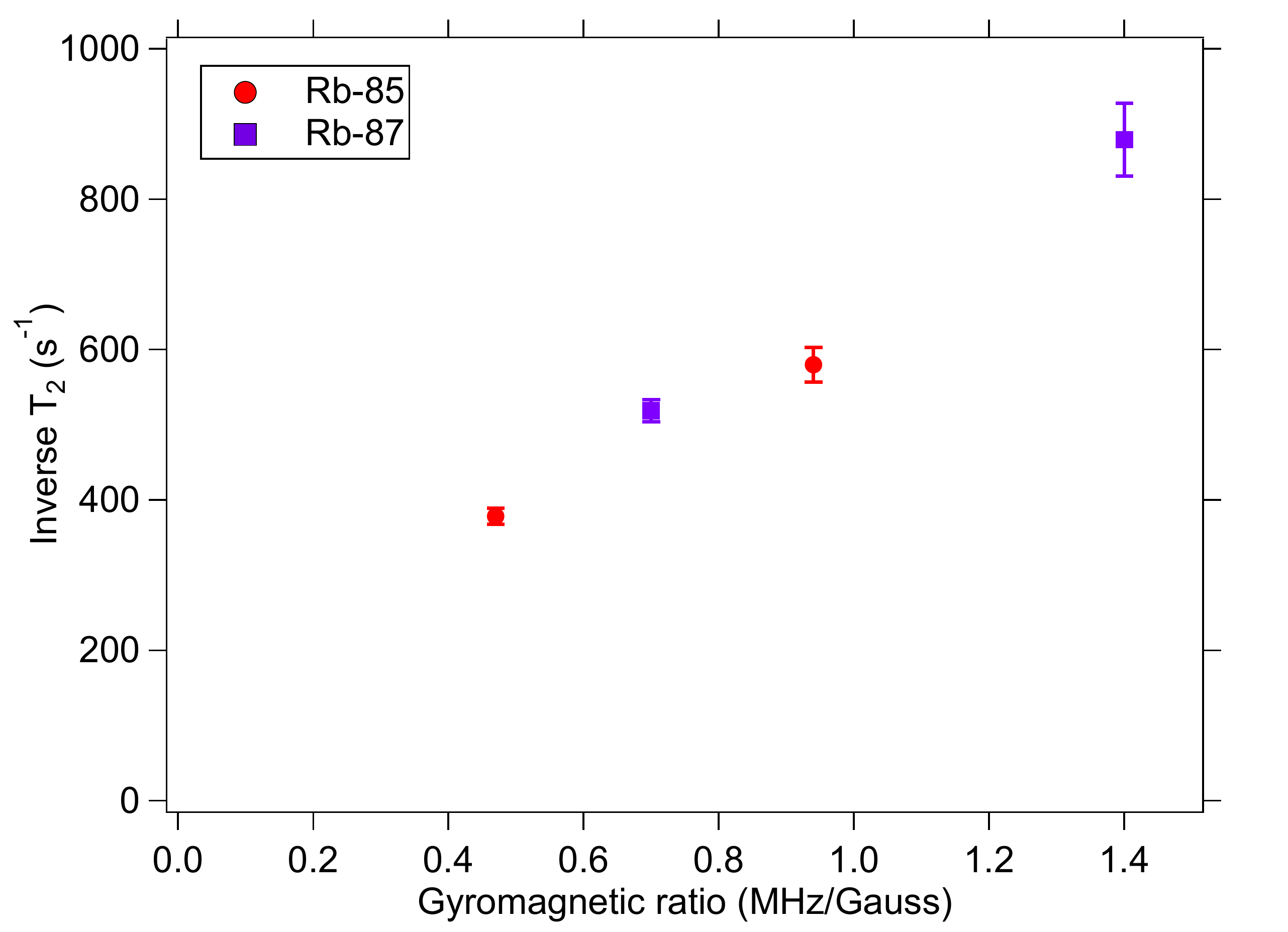}
    \caption{ 
Hahn-echo data for the $m_F = 0,-1$  and  $m_F =+1,-1$ superposition states of $^{85}$Rb and $^{87}$Rb.  The decoherence rate $1/\mathrm{T}_2$ is plotted against the gyromagnetic ratio for each superposition state for each isotope; the $\Delta m_F = 2$ superpositions have a gyromagnetic ratio twice that of the $\Delta m_F=1$ superpositions.  All measurements were performed with the same sample, with an orthohydrogen fraction of $4 \times 10^{-5}$. 
\label{fig:DifferentSuperpositions}
    }
    \end{center}
\end{figure}

The dependence of T$_2$ on species and superposition is very different than what was previously observed for the ensemble dephasing time T$_2^*$.
First,  $^{133}$Cs has a T$_2^*$ roughly one order of magnitude shorter than $^{85}$Rb \cite{PhysRevA.100.063419}. 
Second, for $^{85}$Rb, the $(+1,-1)$ superposition had a significantly longer T$_2^*$ than the $(0,-1)$ superposition \cite{PhysRevB.100.024106}. Both these observations indicate that the dephasing mechanism limiting T$_2^*$ was primarily electrostatic-like in nature \cite{PhysRevB.100.024106, PhysRevA.100.063419}.

Here, we see the opposite behavior.
First, 
as seen in Fig. \ref{fig:HahnEchoLifetimeOrthoCombo}, rubidium and cesium have comparable Hahn-echo coherence times.
Second, as seen in Fig. \ref{fig:DifferentSuperpositions}, the  $(+1,-1)$ superposition has a shorter spin-echo T$_2$ than the $(0,-1)$ superposition. For the different isotopes of rubidium and the different superpositions explored, the decoherence rate $\frac{1}{\mathrm{T}_2}$ is roughly linear in the magnetic field sensitivity of the superposition state. This indicates that the dominant limit on T$_2$ is magnetic-like in nature. 

The source of this magnetic-like noise in our sample has not yet been  identified. Even though the Hahn echo T$_2$ shows little dependence on the orthohydrogen  density at fractions $\lesssim 10^{-4}$, we cannot conclusively rule out the orthohydrogen as the source of the noise, as its nuclear spin T$_1$ and T$_2^*$ have complicated dependences on the orthohydrogen density \cite{buzerak1977nmr, washburn1983pulsed, reif1953nuclear}. It is also possible that other, unknown magnetic impurities  introduced into our parahydrogen matrix during deposition  are limiting the coherence time.
{\color{black}
One such candidate is the HD molecules naturally present in hydrogen. To test the role of HD, we increased the HD fraction in the source gas. Measurements of the resulting samples suggest that  HD impurities are not the dominant limitation on T$_2$, but the measurements were complicated by the observation that HD is preferentially trapped by our ortho-para catalyst, resulting in lower HD fractions in the solid than in the source gas \cite{lorenz2007infrared}. }


We can achieve longer coherence times --- and further learn about the nature of the magnetic-like fluctuations that limit the coherence  --- with Carr-Purcell sequences \cite{PhysRev.94.630}. A schematic of the sequence is shown in Fig. \ref{fig:RawDataAndPulseSequences}.
Applying a standard Carr-Purcell sequence --- in which all $\pi$ pulses are in phase --- resulted in the loss of signal after a small number of pulses ($\lesssim 10$). We attribute this to inaccuracies in $\pi$ pulses which build over successive pulses.
To reduce the problems introduced by imperfect pulses, we use the alternating-phase Carr-Purcell sequence (APCP): the phase of every other $\pi$ pulse was shifted by $180^{\circ}$  to minimize   error accumulation \cite{slichter1990}. 

T$_2$ was measured by two methods: the first  is as previously described and  shown in Fig. \ref{fig:RawDataAndPulseSequences}. 
In the second method we simply monitor the polarization signal as a function of time during the APCP sequence.  Because each $\pi$ pulse rotates the spins through the pole of the Bloch sphere, we are able to effectively measure the readout amplitude at the time of each APCP pulse, allowing for much more rapid data acquisition \cite{PhysRev.94.630}. 
Both methods gave identical results for T$_2$ to within our experimental error. 

\begin{figure}[ht]
    \begin{center}
    \includegraphics[width=\linewidth]{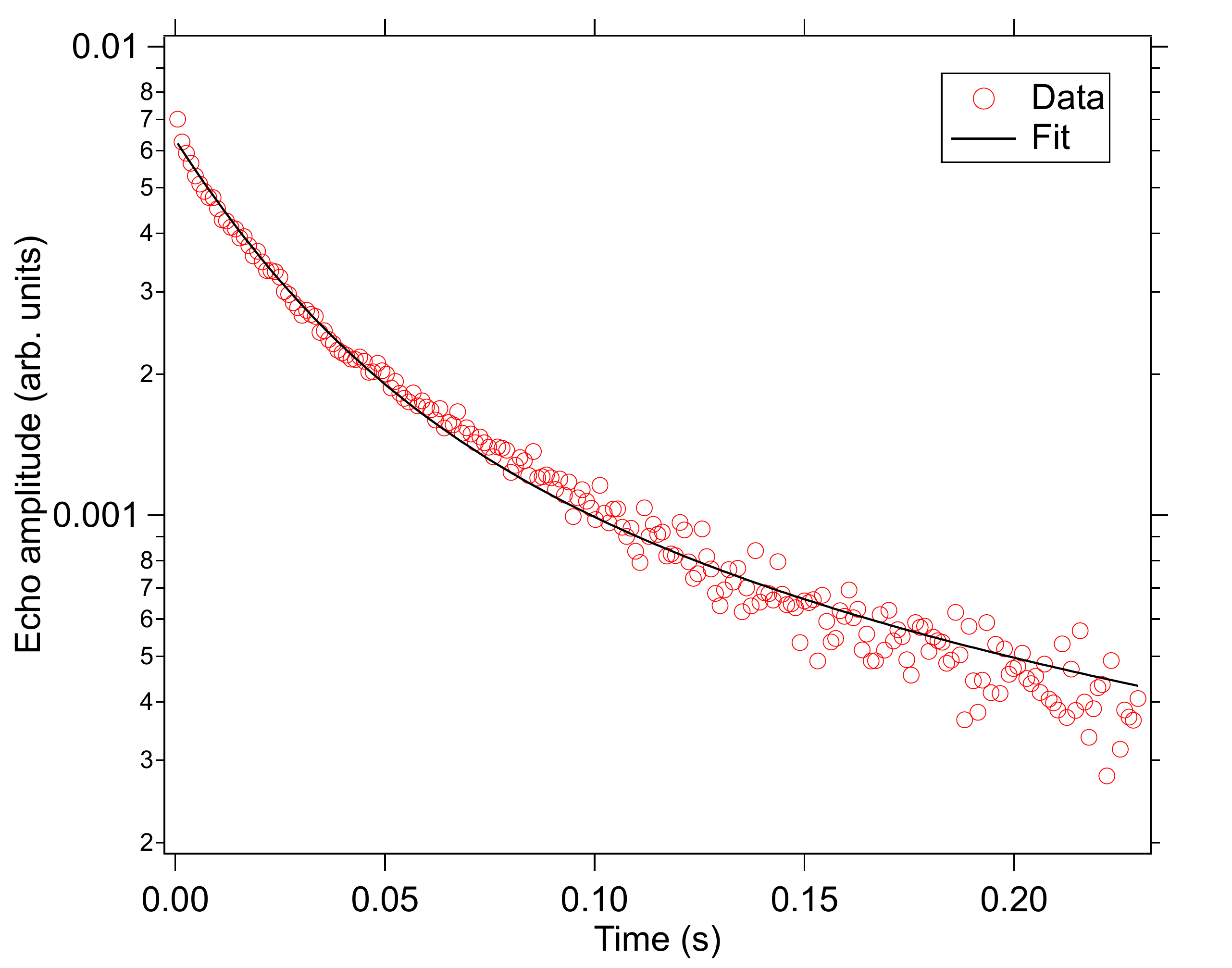}
    \caption{ 
Data taken for Rb atoms with an alternating-phase Carr-Purcell sequence, as discussed in the text, taken at a magnetic field of 45~G, with a 13.25~$\mu$s delay between $\pi$-pulses.
\label{fig:APCPraw}
    }
    \end{center}
\end{figure}

Typical data is shown in Fig. \ref{fig:APCPraw}. The coherence time is significantly longer than what was observed for Hahn echo. However, we note that the decay is poorly described by an exponential (which would appear as a straight line on the log-linear scale of Fig. \ref{fig:APCPraw}).
This is not surprising: we expect an inhomogenous distribution of trapping sites in the sample and consequently a distribution of decoherence rates \cite{upadhyay2016longitudinal}. We model this as a distribution of exponential decay curves; for simplicity we assume a flat distribution of decay rates from zero to some maximum rate. The resulting function is fit to the data (as shown in Fig. \ref{fig:APCPraw}) to determine that maximum rate.  In Fig. \ref{fig:APCPraw}, we see some slight discrepancies between the model at very short times and at very long times. The short time discrepancy is likely due to a long tail of decay rates (missed by our model's sharp cutoff); the long time discrepancy indicates that the distribution does not actually remain constant as the decay rate goes to zero. In the remainder of the paper, we take T$_2$ to be the inverse of the average decay rate.

\begin{figure}[ht]
    \begin{center}
    \includegraphics[width=\linewidth]{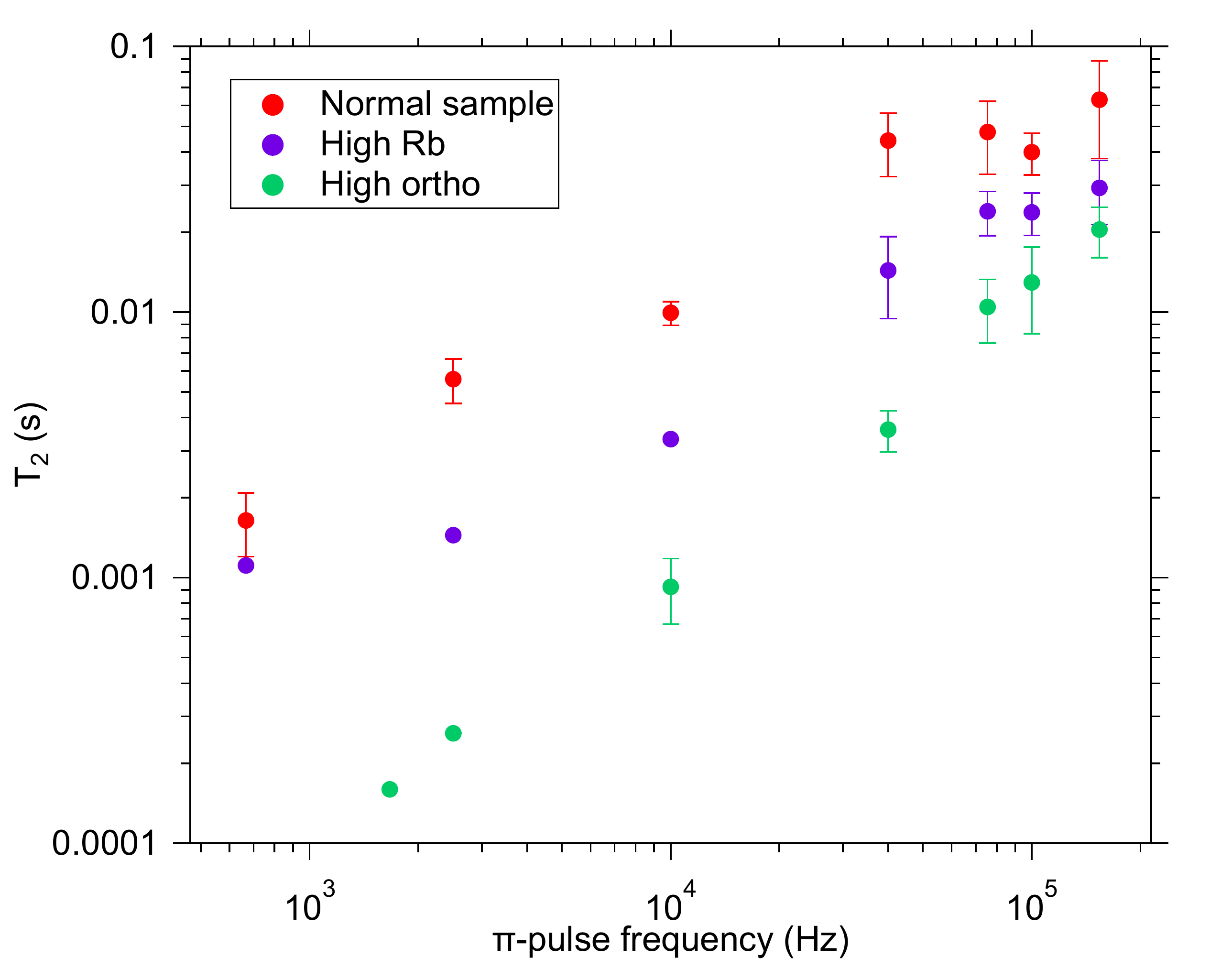}
    \caption{ 
Measured APCP T$_2$ vs $\pi$-pulse repetition rate for different sample conditions. All measurements are for the $F=3$, $m_F = 0, -1$ superposition of $^{85}$Rb. 
The ``normal'' samples have orthohydrogen fractions in the range of $3 \times 10^{-5}$ to $5 \times 10^{-5}$
and rubidium densities from  $5 \times 10^{16}$~cm$^{-3}$ to $1 \times 10^{17}$~cm$^{-3}$.
The ``high ortho'' sample has an orthohydrogen fraction of $1.3 \times 10^{-3}$.
The ``high Rb'' sample has a total rubidium density of $4 \times 10^{17}$~cm$^{-3}$.
The data shown is an average of measurements  from multiple samples, each measured over multiple days. The error bars represent the standard deviations of those measurements (where available; the points missing error bars are expected to have comparable fractional variations). The variation is due to both sample reproducibility and to changes that occur over time, as discussed in the text.
\label{fig:APCP_T2}
    }
    \end{center}
\end{figure}

 The APCP T$_2$ shows a strong dependence on the time delay $\tau$ between the $\pi$ pulses, as seen in Figure \ref{fig:APCP_T2}.
T$_2$ increases with increasing APCP frequency up to the maximum frequency we were able to explore (limited by the duration of our $\pi$ pulses).
During APCP, the superposition is most sensitive to perturbations at a frequency of $\frac{1}{2\tau}$ (using the notation of Fig. \ref{fig:RawDataAndPulseSequences})  and  harmonics \cite{RevModPhys.89.035002}.
The data of Fig. \ref{fig:APCP_T2} indicates that the stochastic magnetic-like fluctuations limiting the Hahn-echo T$_2$ are primarily at  frequencies $\lesssim 1$~kHz. 
Whether longer T$_2$ times could be obtained at even higher APCP frequencies is an open question.

Figure \ref{fig:APCP_T2} shows the measured coherence times for both our highest-purity samples and for samples with elevated rubidium and orthohydrogen densities. 
These lower-purity samples show a measurable decrease in T$_2$.
We model the decoherence at the highest $\pi$-pulse repetition rate, under the assumption that the decoherence rate is linear in both rubidium density and orthohydrogen fraction. The data from the impure crystals  indicates that a significant fraction --- but not all --- of the decoherence in our highest-purity samples is from the rubidium dopants and orthohydrogen impurities. We speculate that the remaining decoherence comes from  the pulse sequence itself.
One source of errors in the pulse sequence is off-resonant coupling out of our two-level system to other Zeeman levels. We observe a reduction in our T$_1$ when we run the APCP pulse sequence at high repetition rates. This effect is more significant (and leads to shorter T$_2$'s) at lower magnetic fields, where the frequency splitting between different $m_F$ transitions is smaller.

For all the samples probed, T$_2$ measurements taken the first day after sample growth are consistently shorter than on subsequent days, and T$_2$  is often observed to continue to slowly increase on a timescale of weeks.
We speculate this is due to the conversion of orthohydrogen to parahydrogen inside our matrix after the sample is grown. Ortho-para conversion in the solid phase has been observed, but under the conditions of our experiment the timescale for conversion in an undoped crystal is much too long to play a significant role  \cite{schmidt1974diffusion, RevModPhys.52.393, shevtsov2000quantum}. Paramagnetic impurities --- such as the rubidium atoms themselves --- are also known to act as a catalyst for ortho-para conversion. However, at the rubidium densities employed in this work, one would expect negligible catalysis of the bulk on the timescale of days \cite{shevtsov2000quantum}. Consistent with this expectation, we see no spectroscopic signs of a significant decrease in the average orthohydrogen fraction after the sample is grown. We speculate that the rubidium atoms are converting some of their nearest-neighbor orthohydrogen molecules (which would be precisely those orthohydrogen molecules that play the most important role in limiting T$_2$), causing  the orthohydrogen fraction in the \emph{local} environment of each rubidium atom to decrease over time.

The long coherence times demonstrated under the APCP protocol make rubidium atoms in parahydrogen very promising for AC magnetic field sensing (at a frequency chosen by the APCP sequence). If detection techniques allow one to efficiently measure single atoms \cite{chambers2019imaging},  a single-atom quantum sensor could be developed. 
This would enable single-molecule NMR experiments \cite{taylor2008high}. Single nitrogen vacancy (NV) sensors in solid diamond have already demonstrated NMR detection of nearby single $^{13}$C nuclear spins within the diamond \cite{zhao2012sensing, kolkowitz2012sensing, taminiau2012detection, abobeih2018one}.  However, the detection of molecules is more difficult: without a method to implant  molecules of interest inside the bulk diamond, the molecules must instead be attached to the surface. 
Unfortunately, the surface is associated with magnetic field noise and significantly reduced NV coherence times \cite{myers2014probing, kim2015decoherence, de2017tailoring, PhysRevX.9.031052, myers2017double, PhysRevX.9.031052}. Parahydrogen, on the other hand, allows for  gentle introduction of molecular species into the bulk during sample growth \cite{Momose1998, yoshioka2006infrared, tam:1926}. 

{\color{black}

We propose that rubidium could be used to make single-molecule NMR measurements of nearby molecules co-trapped within the parahydrogen matrix. 
At a bias magnetic field of 110~G (as was used for the data in Fig. \ref{fig:APCP_T2}), the precession frequencies for $^{13}$C and $^{1}$H would be $1 \times 10^5$~Hz and $5 \times 10^5$~Hz respectively.
%
%
Following the protocol of Ref. \cite{zhao2012sensing}, one can detect nuclear spins using a APCP sequence with $\pi$-pulses at twice the precession frequency. 
This is slightly outside the pulse frequency range explored in this work, but we expect it is straightforward to achieve with higher-power RF electronics.
Assuming we are able to efficiently detect the spin state of a \emph{single} rubidium atom, if we scale the results of Ref. \cite{zhao2012sensing} using the  coherence times measured in this work, we would expect to be able to sense a single proton at a distance of 10~nm within 1~s.
%
%
%
%

With the addition of field gradients, this could be extended to perform magnetic resonance imaging of the structure of single molecules, as was  previously proposed for NV centers \cite{PhysRevX.5.011001, staudacher2013nuclear, mamin2013nanoscale, sushkov2014magnetic}. Nuclear spin imaging at the single-nucleus level would be of tremendous value for understanding biochemistry and for applications in medicine and drug development.
In future work, we hope to move from the ensemble measurements presented here to the single-atom measurements needed for single-molecule NMR and MRI.
}

Our longest measured APCP T$_2$ time, for our best sample, was 0.1~s.
This is over an order-of-magnitude longer than has been achieved with near-surface NV centers to date {\color{black} \cite{myers2014probing, kim2015decoherence, de2017tailoring, PhysRevX.9.031052, myers2017double, PhysRevX.9.031052}}. 
In future work it may be possible to achieve longer spin coherence times with the use of more sophisticated dynamical decoupling pulse sequences \cite{barry2019sensitivity} and with the growth of higher-purity samples.  It may also be possible to achieve greater magnetic field sensitivity with nonclassical superposition states \cite{PhysRevB.100.024106}.

\section*{Acknowledgements}
This material is based upon work supported by the National Science Foundation under Grants No. PHY-1607072 and PHY-1912425.
We gratefully acknowledge helpful conversations with Amar C. Vutha.

\bibliography{parahydrogen_alkali_T2.bib}

\end{document}